\documentclass[a4paper,11pt]{article}
\usepackage[dvipdfm]{graphicx}
\usepackage{natbib}
\begin{document}
\begin{titlepage}
\begin{flushright}
\end{flushright}
\vfill
\begin{center}
{\Large \bf
  Combined effect of successive competition periods
  on population dynamics
}
\vfill
 {\bf Masahiro Anazawa}%
 \footnote[1]{Tel.:+81-22-305-3931; Fax:+81-22-305-3901.
E-mail address: anazawa@tohtech.ac.jp.}
\vfill
{\itshape
  Department of Environmental Information Engineering, \\
  Tohoku Institute of Technology, Sendai 982-8577, Japan \\
}
\vfill
\end{center}

\vfill
\begin{abstract}
This study investigates the effect of competition between individuals on
population dynamics when they compete for different resources during
different seasons or during different growth stages.  Individuals are
assumed to compete for a single resource during each of these periods
according to one of the following competition types: scramble, contest,
or an intermediate between the two.  The effect of two successive
competition periods is determined to be expressed by simple relations on
products of two ``transition matrices'' for various sets of competition
types for the two periods.  In particular, for the scramble and contest
competition combination, results vary widely depending on the order of
the two competition types.  Furthermore, the stability properties of
derived population models as well as the effect of more than two
successive competition periods are discussed.
\end{abstract}
\vfill
\vfill
\end{titlepage}


\section{Introduction}
\label{sect:introduction}
Discrete-time population models based on a difference equation have been
widely employed for modeling population dynamics in species with
seasonal reproduction, but in many cases, these models have been
introduced as phenomenological models.  Since population dynamics must,
in principle, result from behaviors of individuals comprising a
population, it is important to clarify the relationships between
processes on both individual and population levels \citep{hasmay1985,
Lom1988}.  An effective approach is to derive population models from
first principles, and studies of this kind have recently advanced a lot.
First-principles derivations are broadly classified into continuous-time
approaches \citep[e.g.,][]{gur1998, welstr1998, thi2003, gerkis2004,
eskger2007, eskpar2007, eskpar2010} and discrete-time approaches
\citep[e.g.,][]{roy1992, sumbro2001, johsum2003, brasum2005a,
brasum2005b, brasum2006, sen2007, ana2009, ana2010}.  There are also
studies employing individual-based simulations
\citep[e.g.,][]{johber2008}.  Extending the study of \cite{ana2010},
which is based on a discrete-time approach, this study investigates the
effect of competition on population dynamics when individuals compete
for different resources during different seasons or during different
growth stages.

It is well known that there are two contrasting types of competition
between individuals: scramble and contest \citep{nic1954, has1975}.  In
scramble, resources are considered to be partitioned evenly, while in
contest, resources are monopolized by a few competitively superior
individuals.  Extending the study of \cite{brasum2005b}, \cite{ana2010}
derived a population model for a competition type intermediate between
scramble and contest through the consideration of resource partitioning
and spatial distribution of individuals, and showed that this model
includes various classical population models as special cases in various
limits in terms of two parameters.  However, this derivation dealt only
with cases in which individuals compete for the same resource throughout
the competition period, and not with cases in which individuals compete
for different resources simultaneously or cases in which variation of
resource and competition type depend on seasons or on growth stages of
individuals.

Extending the study of \cite{ana2010}, this study investigates the
effect of competition in situations wherein the overall competition
period consists of different competition periods.  Individuals compete
for a single resource within each period, but the types of resource and
competition vary between competition periods.  The effect of two
successive competition periods is expressed by simple relations on
products of two ``transition matrices.''  The results depend on the
competition types in the two periods, and in particular, exhibit a great
deal of contrast for the scramble and contest competition combination,
depending on the order of the two competition types.  Furthermore, the
stability properties of derived population models as well as the effect
of more than two successive competition periods are discussed.

\section{The basic framework}
\label{sect:basic}

\subsection{Site-based framework}
\label{subsect:site-based}
\begin{figure}[htbp]
\centering \includegraphics[scale=0.8,clip]{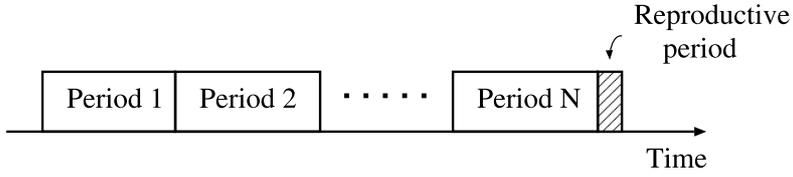}
\caption{
The structure of the overall competition period.  The overall
competition period consists of $N$ distinct competition periods, where
individuals compete for different resources during these different
periods.
}
\label{fig:periods}
\end{figure}

A site-based framework, described below, is assumed as the basic
framework for considering the derivation of population models
\citep{sumbro2001, johsum2003, brasum2005b, sen2007, ana2009, ana2010}.
This is a modified version of the framework used in the study of
\cite{ana2010}, designed to deal with multiple competition periods.
Consider a habitat consisting of $n$ discrete resource sites or patches
over which all individuals of a population of size $x_t$ in generation
$t$ are distributed.  Not moving to other sites, each individual is
assumed to compete for resources with other individuals at the same
site.  As will be described in Section~\ref{subsect:competition}, the
overall period in which individuals compete for resources consists of
multiple distinct competition periods, wherein the types of resource and
competition depend on these competition periods
(Fig.~\ref{fig:periods}).  Individuals that procure sufficient resources
to grow and survive until the reproductive period, reproduce, and all
parents die after the reproductive period.  Offspring emerging from each
site then disperse and are distributed randomly over the sites again,
forming a population in the next generation.

In this situation, the expected population size in generation $t+1$ can
be written as follows:
\begin{equation}
\label{eq:next_generation}
 x_{t+1}=n \sum_{k=1}^{\infty} p_k(x_t) \, \phi(k).
\end{equation}
Here, $\phi(k)$, referred to as the interaction function, denotes the
expected number of offspring emerging from a site containing $k$
individuals, and its specific forms are determined through the
consideration of competition for resources between individuals in a
site.  On the other hand, $p_k(x_t)$, a function of $x_t$, denotes the
probability of finding $k$ individuals at a given site at the beginning
of the overall competition period.  Eq.~(\ref{eq:next_generation})
connects population dynamics at two different spatial scales: the
population dynamics within each site, represented by $\phi(k)$; and
those over the entire habitat consisting of all sites. In this sense,
the site-based framework is closely related to scale transition theory
\citep{che1998a, che1998b, chedon2005}.

As described in \cite{ana2010}, for a situation in which the
distribution of individuals shows aggregated patterns, we assume $p_k$
to be the following negative binomial distribution:
\begin{equation}
\label{eq:negative_binomial}
  p_k = \frac{\Gamma(k+\lambda)}{\Gamma(\lambda)\Gamma(k+1)}
        \left(\frac{x_t}{\lambda n}\right)^k 
        \left(1+\frac{x_t}{\lambda n}\right)^{-k-\lambda},
\end{equation}
where $\lambda$ is a positive parameter, with $1/\lambda$ representing
an index of the degree of aggregation.  This distribution includes the
following Poisson distribution as a special case in the limit as
$\lambda\to\infty$:
\begin{equation}
\label{eq:Poisson}
  p_k = \frac{1}{k!}\left(\frac{x_t}{n}\right)^k e^{-x_{t}/n},
\end{equation}
which corresponds to a situation in which individuals are distributed
completely at random.  Here, the number $n$ of the resource sites is
assumed to be sufficiently large.

\subsection{Competition for resources}
\label{subsect:competition}
We now consider the competition for resources within each site more
concretely.  Consider that the entire competition period, from the time
when individuals enter the site to the beginning of the reproductive
period, consists of $N$ distinct competition periods
(Fig.~\ref{fig:periods}).  Although individuals compete for a single
resource during each period, the types of resource and competition
depend on the competition period.  Letting resource $i$ represent the
resource for which individuals compete in period $i$, each individual is
assumed to require an amount $s_i$ of resource $i$ in order to survive
until the end of period $i$ and advance to the next period; if an
individual alive at the beginning of period $i$ procures the resource
amount $s_i$ within this period, it advances to the next period, if not,
it dies within period $i$.  Individuals alive at the end of period $N$
advance to the reproductive period and reproduce.  Furthermore, the
total amount $R_i$ of resource $i$ contained in each site is assumed to
follow an exponential distribution with the following probability
density function:
\begin{equation}
\label{eq:qR}
 q_i(R_i)= e^{-R_i/\bar{R}_i} / \bar{R}_i,
\end{equation}
where $\bar{R}_i$ denotes the mean of $R_i$.  This corresponds to a
situation in which the total amount $n \bar{R}_i$ of the resource is
partitioned among the $n$ sites completely at random.  Distribution of
each resource is assumed to be independent.

In period $i$, the amount $R_i$ of resource $i$ in each site is
partitioned between the individuals alive at the site, and the type of
resource partitioning is assumed to be one of the following three types,
as described in \cite{ana2009, ana2010}:
\begin{itemize}
\item[(a)]%
{\bf Scramble competition} \\
The resource $R_i$ is evenly partitioned between the individuals at the
site.
\item[(b)]%
{\bf Contest competition} \\ 
The resource $R_i$ is partitioned between the individuals in the order
of competitive ability, i.e., the most competitive one first takes the
amount $s_i$, and then, if the amount of the remaining resource is $\ge
s_i$, the next most competitive one takes the amount $s_i$, and this
process repeats until $R_i$ is depleted or all individuals procure
$s_i$.
\item[(c)] %
{\bf Intermediate competition} \\ 
This is an intermediate between the above two types.  First, a certain
amount $\hat{s}_i\;(< s_i)$ is equally distributed to each individual at the
site, and then the remaining resource is partitioned in the order of
competitive ability, i.e., the most competitive one first tries to
take the amount $s_i-\hat{s}_i$ so as to obtain a total amount
$s_i$. Thereafter, if the amount of the remaining resource is 
$\ge s_i-\hat{s}_i$, the next most competitive one takes
$s_i-\hat{s}_i$, and this process repeats until $R_i$ is depleted or 
all individuals procure $s_i$.
\end{itemize}
Table~\ref{table:N=1} presents the interaction functions and population
models for the three competition types described above when $N=1$
\citep{ana2009, ana2010}.  The six functions defined here are used
throughout this study.  In this table, $\alpha=e^{-s/\bar{R}}$ and
$\hat{\alpha}=e^{-\hat{s}/\bar{R}}$, and $\beta$ is defined by
\begin{equation}
 \label{eq:beta}
 \beta = \frac{1-\hat{\alpha}}{1-\alpha},
\end{equation}
representing the degree of deviation from (ideal) contest competition in
the case of intermediate competition ($0<\beta<1$).  Intermediate
competition approaches contest type in the limit as $\beta\to 0$
($\hat{s}\to 0$) and scramble type in the limit as $\beta\to 1$
($\hat{s}\to s$).  As described in \cite{ana2010}, the population model
$x_{t+1}=f_I(x_t; \alpha, \beta, \lambda)$ for intermediate competition
includes the following classical population models as special cases in
various limits in terms of $\beta$ and $\lambda$: the Ricker model
\citep{ric1954}, the Hassell model \citep{has1975, haslaw1976,
dej1979}, the Skellam model \citep{ske1951}, the Beverton-Holt model
\citep{bevhol1957}, and the Br{\" a}nnstr{\" o}m-Sumpter model
\citep{brasum2005b}.


\section{Combined effect of competition periods}
\label{sect:general}

\subsection{Transition matrix}
We now consider how to derive specific forms of $\phi(k)$ for the
overall competition period consisting of $N$ distinct periods.  Here, we
should note that the number of individuals alive at a site can vary with
time due to competition.  Let $T_{k, m}$, henceforth referred to as the
transition probability, be the probability that the number of the
individuals alive at a given site changes from $k$ at the beginning to
$m$ at the end of the overall competition period; and let $T$, referred
to as the transition matrix, be an infinite size matrix with its $(k,
m)$ component given by $T_{k, m}$.  The transition probability $T_{k,
m}$ can be written as follows:
\begin{equation}
 \label{eq:transition}
 T_{k, m}=\sum_{l_1=1}^{k} \sum_{l_2=1}^{k} \cdots
 \sum_{l_{N-1}=1}^{k}
 T^{(1)}_{k,l_1}\,T^{(2)}_{l_1, l_2}\cdots
 T^{(N)}_{l_{N-1}, m},
\end{equation}
where $T^{(i)}_{k,m}$ denotes the transition probability for period $i$,
i.e., the probability of the number of individuals alive at a given site
changing from $k$ at the beginning to $m$ at the end of period $i$.
Eq.~(\ref{eq:transition}) can be written concisely in a matrix form as
follows:
\begin{equation}
 \label{eq:transition2}
 T = T^{(1)} \, T^{(2)}\cdots T^{(N)}.
\end{equation}
The interaction function $\phi(k)$ is determined from the transition
matrix by using the relation
\begin{equation}
 \label{eq:phi_T}
 \phi(k)= b' \sum_{m=1}^{k} T_{k, m}\,m,
\end{equation}
where $b'$ is a positive parameter representing the average number of
offspring produced by an individual alive in the reproductive period
multiplied by the survivorship of the offspring until they enter their
habitation sites.

\subsection{Product of two transition matrices}
Next, we determine a product of two transition matrices, which
corresponds to the $N=2$ case of Eq.~(\ref{eq:transition}).  As will be
shown later, products of more than two transition matrices can be
determined by recursive use of this result.  Given that the competition
within period $i$ is the intermediate type specified by
$(\hat{\alpha}_i,\alpha_i) =(e^{-\hat{s}_i/\bar{R}_i},
e^{-s_i/\bar{R}_i})$, the transition probabilities for this period are
written as follows:
\begin{equation}
 \label{eq:T_i0}
 T_I(\hat{\alpha}_i, \alpha_i)_{k, m} 
 = \phi_I(\hat{\alpha}_i, \alpha_i)_{k, m} 
 - \phi_I(\hat{\alpha}_i, \alpha_i)_{k, m+1},
\end{equation}
where $\phi_I(\hat{\alpha}_i, \alpha_i)_{k, m}$ denotes the probability
that at least $m$ individuals survive until the end of period $i$ at a
given site where $k$ individuals were alive at the beginning.  For the
$m$-th individual, in the order of competitive ability, to be able to
obtain the amount $s_i$ of resource so as to survive till the end of the
period, $R_i \ge \hat{s}_i k +(s_i-\hat{s}_i)m$ must be satisfied, and
hence,
\begin{equation}
 \label{eq:phi_i0}
 \phi_I(\hat{\alpha}_i, \alpha_i)_{k, m} 
 =\mbox{Prob}[R_i \ge \hat{s}_i k +(s_i-\hat{s}_i)m] \, I_{k\ge m},
\end{equation}
where $I_{k\ge m}$ is defined by
\[
 I_{k\geq m} = \biggl\{
    \begin{array}{ll}
     1 & \quad (\mbox{for $k \geq m$}), \\
     0 & \quad (\mbox{otherwise}).
    \end{array}\biggr. 
\]
Determining the probability in the right-hand side of
Eq.~(\ref{eq:phi_i0}) from distribution (\ref{eq:qR}), and combining the
obtained result with Eq.~(\ref{eq:T_i0}), yields the following
transition probabilities:
\begin{equation}
 \label{eq:psi_i}
 T_I(\hat{\alpha}_i, \alpha_i)_{k, m} 
 =\cases{\hat{\alpha}_i^k (\alpha_i/\hat{\alpha}_i)^m 
 (1-\alpha_i/\hat{\alpha}_i) & 
 \quad $(1\le m \le k-1)$, \cr
 \alpha_i^k & \quad $(m=k)$, \cr
 0 & \quad $(m\ge k+1)$.}
\end{equation}
Substituting the above probabilities with $i=1, 2$ into
Eq.~(\ref{eq:transition}), and performing a summation, renders the
following relation (for details, see Appendix A):
\begin{eqnarray}
 \label{eq:transition_i_result}
 \lefteqn{T_I(\hat{\alpha}_1, \alpha_1) \, T_I(\hat{\alpha}_2, \alpha_2)}
 \nonumber \\
  &=& K \,T_I(\hat{\alpha}_1, \alpha_1 \alpha_2) 
  + (1-K)\, T_I(\alpha_1 \hat{\alpha}_2, \alpha_1 \alpha_2),
\end{eqnarray}
where $K$ is a constant ($0<K<1$) defined by
\begin{equation}
 K=\frac{\hat{\alpha}_1-\alpha_1}
 {\hat{\alpha}_1-\alpha_1 \hat{\alpha}_2}.
\end{equation}
This result indicates that the product of two transition matrices of
intermediate competition is equivalent to a linear combination of two
other transition matrices of the intermediate type.  In other words, the
effect of overall competition is a linear interpolation of the effects
of two distinct intermediate competitions.

Also, combining Eq.~(\ref{eq:transition_i_result}) and
Eq.~(\ref{eq:phi_T}) gives
\begin{equation}
 \label{eq:phi_i_result}
 \phi(k)= K\, \phi_I(k\,; \hat{\alpha}_1, \alpha_1 \alpha_2) 
 + (1-K)\, \phi_I(k\,;\alpha_1\hat{\alpha}_2, \alpha_1 \alpha_2),
\end{equation}
where $\phi_I$ is the interaction function for intermediate competition,
defined in Table \ref{table:N=1}.  Similarly, the corresponding
population model is given by
\begin{equation}
 \label{eq:pop_i_result}
 x_{t+1}= K\, f_I(x_t\,; \alpha_1 \alpha_2, \beta_1, \lambda)
 + (1-K)\, f_I(x_t\,; \alpha_1 \alpha_2, \beta_2, \lambda),
\end{equation}
where $\beta_1$ and $\beta_2$ are given, from Eq.~(\ref{eq:beta}), by
\begin{equation}
 \beta_1 = \frac{1-\hat{\alpha}_1}{1-\alpha_1 \alpha_2},
\end{equation}
\begin{equation}
 \beta_2 = \frac{1-\alpha_1 \hat{\alpha}_2}{1-\alpha_1 \alpha_2}.
\end{equation}
Eq.~(\ref{eq:pop_i_result}) is a linear combination of two population
models for intermediate competition with different values of $\beta$.
Note that these values of $\beta$ satisfy $\beta_1 < \beta_2$, since
$0<\alpha_i<\hat{\alpha}_i <1$ gives $\alpha_1 \hat{\alpha}_2 < \alpha_1
< \hat{\alpha}_1$, and then $1- \hat{\alpha}_1 < 1- \alpha_1
\hat{\alpha}_2 $.

\subsection{Cases of scramble and contest competition}
Since the intermediate competition specified by $(\hat{\alpha}_i,
\alpha_i)$ approaches scramble in the limit as $\hat{\alpha}_i\to
\alpha_i$, and contest in the limit as $\hat{\alpha}_i\to 1$,
considering such limits in Eq.~(\ref{eq:transition_i_result}) gives four
other relations involving the transition matrices for scramble and
contest, as shown in Table~\ref{table:products_of_matrices}.  In this
table, $T_S(\alpha_i)$ and $T_C(\alpha_i)$ denote the transition
matrices for scramble and contest competitions, respectively.
Furthermore, these four relations yield similar relations about
interaction functions, which then give the corresponding population
models for various combinations of scramble and contest competitions
shown in Table~\ref{table:pop_models}.

In Table~\ref{table:products_of_matrices}, it is interesting to note
that the product of two transition matrices of the same type of either
scramble or contest competition is equivalent to a single transition
matrix of the same type.  Further, the products of a matrix for scramble
and that for contest are different, depending on the combination order
of the competition types (discussed further in
Section~\ref{subsect:sc_vs_cs}).  Also, note that the set of relations
in Eqs.~(\ref{eq:tra_ss})--(\ref{eq:tra_cs}) is mathematically
equivalent to the original relation elaborated in
Eq.~(\ref{eq:transition_i_result}), since the latter can be obtained
reversely from the former (see Appendix B).

\newcounter{CNT}
\setcounter{CNT}{\value{equation}}
\addtocounter{equation}{11}

\subsection{Emergence of a linear combination}
The relation in Eq.~(\ref{eq:tra_cs}) in
Table~\ref{table:products_of_matrices} indicates that the effect of
contest followed by scramble is equivalent to a linear combination of
the effects of another contest and another scramble.  This relation was
derived in a non-intuitive manner, but it is possible to intuitively
understand why this linear combination emerges when $s_i/\bar{R}_i \ll
1$, as described below.

For simplicity, we consider the corresponding relation about interaction
functions instead of Eq.~(\ref{eq:tra_cs})
\begin{equation}
      \phi(k)=\frac{1-\alpha_1}{1-\alpha_1 \alpha_2} \phi_C(k\,; 
      \alpha_1 \alpha_2)
      + \frac{\alpha_1(1-\alpha_2)}{1-\alpha_1 \alpha_2} 
      \phi_S(k\,; \alpha_1 \alpha_2).
 \label{eq:phi_cs_1}
\end{equation}
When $s_i/\bar{R}_i \ll 1$, this equation can be approximately written
as
\begin{eqnarray}
 \phi(k)&\simeq&
 \frac{s_1/\bar{R}_1}{s_1/\bar{R}_1 + s_2/\bar{R}_2}\,
 \phi_C(k\,; \alpha_1 \alpha_2) \nonumber \\
 &&+\frac{s_2/\bar{R}_2}{s_1/\bar{R}_1 + s_2/\bar{R}_2}\,
 \phi_S(k\,; \alpha_1 \alpha_2),
 \label{eq:phi_cs_2}
\end{eqnarray}
which can be derived as follows.  Consider a site with $R_1$, $R_2$
resources, and $k$ individuals alive at the beginning.  The individuals
compete for $R_1$ in period $1$, and for $R_2$ in period $2$.  The
maximum number of individuals that can live on $R_i$ in period $i$ is
given by the maximum integer not exceeding $R_i/s_i$, which is
approximately $R_i/s_i $ when $s_i/\bar{R}_i \ll 1$.  This indicates
that the overall competition is dominated by the competition with a
smaller value of $R_i/s_i$.  In other words, it is dominated by the
competition for $R_1$ (i.e., contest competition) when $R_1/s_1 <
R_2/s_2$, and by the competition for $R_2$ (i.e., scramble competition)
when $R_1/s_1 > R_2/s_2$.  Hence, the interaction function can be
written as follows:
\begin{eqnarray}
 \phi(k)&\simeq&
  b' \sum_{m=1}^{k} \mbox{Prob}[m \le R_1/s_1 < R_2/s_2]
  \nonumber \\
  && +\; b' \, \mbox{Prob}[k \le R_2/s_2 < R_1/s_1] \,k.
 \label{eq:phi_cs_3}
\end{eqnarray}
Here, the first term on the right-hand side represents the contribution
from the contest competition, which dominates when $R_1/s_1 < R_2/s_2$,
with $m \le R_1/s_1 < R_2/s_2$ being the condition for the $m$-th
individual, in the order of competitive ability, to be able to obtain
$s_1$ from $R_1$.  The second term represents the contribution from the
scramble competition, which dominates when $R_1/s_1 > R_2/s_2$, with $k
\le R_2/s_2 < R_1/s_1$ being the condition for all individuals to be
able to obtain $s_2$ from $R_2$.  An explicit calculation using the
distribution of $R_i$, Eq.~(\ref{eq:qR}), shows that the first and the
second terms on the right-hand side of Eq.~(\ref{eq:phi_cs_3}) are
equivalent to the first and second terms of Eq.~(\ref{eq:phi_cs_2}),
respectively.  The derivation of Eq.~(\ref{eq:phi_cs_2}) provided here
explains intuitively why a linear combination of scramble and contest
emerges in the $s_i/\bar{R}_i \ll 1$ case.

\subsection{Scramble followed by contest vs. contest followed by
  scramble}
\label{subsect:sc_vs_cs}
Population models (\ref{eq:pop_sc}) and (\ref{eq:pop_cs}) in
Table~\ref{table:pop_models} for the combination of scramble and contest
depend on the order of these types of competition.  In the following, a
comparison between these two models is presented.  Consider $\alpha_s$
and $\alpha_c$ to be the parameters characterizing scramble and contest
competitions, respectively. The model for scramble followed by contest
competition (the SC model) is written from Eq.~(\ref{eq:pop_sc}) as
follows:
\begin{equation}
 \label{eq:pop_sc_}
 x_{t+1} = f_I(x_t\, ; \alpha, \beta, \lambda),
\end{equation}
where $\alpha=\alpha_s \alpha_c$ and $\beta=(1-\alpha_s)/(1-\alpha)$.
On the other hand, the model for contest followed by scramble
competition (the CS model) is written from Eq.~(\ref{eq:pop_cs}) as
follows:
\begin{equation}
 \label{eq:pop_cs_}
 x_{t+1} = \gamma_c \,f_C(x_t\, ; \alpha, \lambda)
         +\gamma_s \,f_S (x_t\, ; \alpha, \lambda),
\end{equation}
where $\gamma_s=\alpha_c(1-\alpha_s)/(1-\alpha)$ and
$\gamma_c=1-\gamma_s$.  Here, both $\beta$ and $\gamma_s$ take values
from $0$ to $1$. Both models approach the model for contest as these
parameters approach $0$, and the model for scramble as they approach
$1$.  In this sense, both $\beta$ and $\gamma_s$ are indices
representing the degree of derivation from the model for (ideal) contest
competition.

\begin{figure}[htbp]
\centering
\includegraphics[scale=1.1,clip]{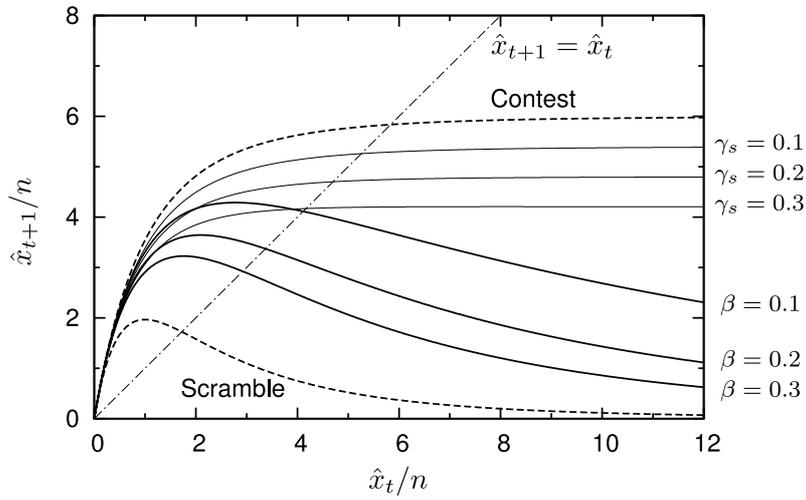}
\caption{
Comparison of reproduction curves of the SC model (scramble followed by
contest, thick solid lines) and the CS model (contest followed by
scramble, thin solid lines) in the case of $b'\alpha=6$ and $\lambda=4$,
where $\hat{x}_t=(1-\alpha)x_t$.  Both $\beta$ and $\gamma_s$ represent
degrees of deviation from (ideal) contest competition.  The dashed lines
represent the reproduction curves for a single period of contest and for
a single period of scramble.
}
\label{fig:curves}
\end{figure}

Although both Eq.~(\ref{eq:pop_sc_}) and Eq.~(\ref{eq:pop_cs_}) can be
regarded as population models intermediate between the models for
contest and scramble, they show very different reproduction curves as
shown in Fig.~\ref{fig:curves}.  When $\hat{x}_t/n \equiv
(1-\alpha)x_t/n$ is sufficiently small, models (\ref{eq:pop_sc_}) and
(\ref{eq:pop_cs_}) can be approximately written as
\begin{equation}
 \hat{x}_{t+1}\simeq b'\alpha\, \hat{x}_t \left\{1-(1+\beta)
 \Bigl(1+\frac{1}{\lambda}\Bigr) \frac{\hat{x}_t}{2n}\right\},
\end{equation}
\begin{equation}
 \hat{x}_{t+1}\simeq b'\alpha\, \hat{x}_t 
 \left\{1-(1+\gamma_s)
 \Bigl(1+\frac{1}{\lambda}\Bigr) \frac{\hat{x}_t}{2n}\right\},
\end{equation}
respectively. This shows that the reproduction curves of the two models
with the same value of $\beta$ and $\gamma_s$ agree with each other for
small values of $\hat{x}_t/n$.  However, as $\hat{x}_t/n$ becomes
larger, the difference between the two curves increases, and
$\hat{x}_{t+1}/n$ approaches $0$ for the SC model, and $\gamma_c
b'\alpha$ for the CS model.  In general, curves of the SC model resemble
that of (ideal) scramble competition and curves of the CS model resemble
that of (ideal) contest competition.  The rationale for such a
difference can be explained as follows.  When the population size is
large, the competition in the first period is severe, thus giving rise
to a large reduction in the number of individuals alive in each
site. This results in a small impact due to competition in the second
period.  In this way, the competition in the first period dominates the
entire population dynamics when the population size is large.

\begin{figure}[htbp]
\centering
\includegraphics[scale=1.4,clip]{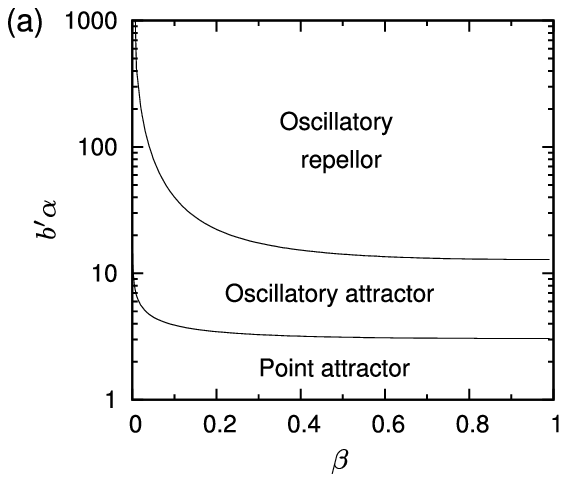}
\\
\vspace{10mm}
\includegraphics[scale=1.4,clip]{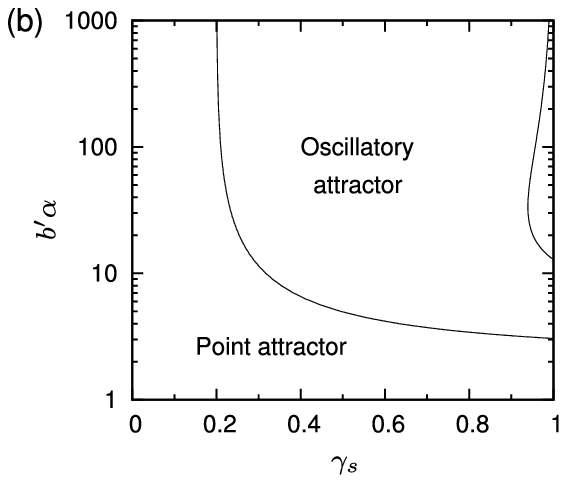}
\caption{
Diagrams showing the properties of the positive equilibrium point of (a)
the SC model, and of (b) the CS model in the case of $\lambda=4$.  In
panel (b), the small region on the right side is the region of the
oscillatory repellor.
}
\label{fig:SC+CS}
\end{figure}

Next, we compare the stability properties of the two population models.
Both models have a positive equilibrium point when $b'\alpha>1$.
Fig.~\ref{fig:SC+CS} illustrates how the stability properties of the
positive equilibrium point depend on $b'\alpha$ and $\beta$ for
$\lambda=4$.  Fig.~\ref{fig:SC+CS}~(a) shows that, for the SC model, the
properties of the equilibrium point exhibit an abrupt change with a
small increase in $\beta$ from $0$, but Fig.~\ref{fig:SC+CS}~(b) shows
that the same is not true for the CS model with a small increase in
$\gamma_s$ from $0$.  Furthermore, as shown in Fig.~\ref{fig:SC+CS}~(b),
the equilibrium point of the CS model is always a point attractor when
$\gamma_s<1/(\lambda+1)$ (see Appendix C).

\subsection{Products of more than two transition matrices}
Products of more than two transition matrices can be determined by the
recursive use of the relation described in
Eq.~(\ref{eq:transition_i_result}).  For example, multiplying both sides
of Eq.~(\ref{eq:transition_i_result}) on the right by
$T_I(\hat{\alpha}_3, \alpha_3)$ gives
\begin{eqnarray}
 \lefteqn{T_I(\hat{\alpha}_1, \alpha_1) \, T_I(\hat{\alpha}_2, \alpha_2)
 \, T_I(\hat{\alpha}_3, \alpha_3)}
 \nonumber \\
  &=& \left\{ K \,T_I(\hat{\alpha}_1, \alpha_1 \alpha_2) 
  + (1-K)\, T_I(\alpha_1 \hat{\alpha}_2, \alpha_1 \alpha_2) \right\}
  \, T_I(\hat{\alpha}_3, \alpha_3) \nonumber \\
  &=&  K \,T_I(\hat{\alpha}_1, \alpha_1 \alpha_2)
  \, T_I(\hat{\alpha}_3, \alpha_3) \nonumber \\
  && + \,\,\, (1-K)\, T_I(\alpha_1 \hat{\alpha}_2, \alpha_1 \alpha_2) 
  \, T_I(\hat{\alpha}_3, \alpha_3).
\end{eqnarray}
Rewriting the right-hand side with the relation
(\ref{eq:transition_i_result}) gives
\begin{eqnarray}
 \label{eq:transition_3i}
 \lefteqn{T_I(\hat{\alpha}_1, \alpha_1) \, T_I(\hat{\alpha}_2, 
 \alpha_2) \, T_I(\hat{\alpha}_3, \alpha_3)}
 \nonumber \\
  &=& K_{1} \,T_I(\hat{\alpha}_1, \alpha_1 \alpha_2 \alpha_3) 
  + K_{2}\, T_I(\alpha_1 \hat{\alpha}_2, \alpha_1 \alpha_2 \alpha_3) 
  \nonumber \\
  && + \,\,\, K_{3}\, T_I(\alpha_1 \alpha_2 \hat{\alpha}_3, 
  \alpha_1 \alpha_2 \alpha_3),
\end{eqnarray}
where
\begin{eqnarray}
 K_{1}&=&\frac{\hat{\alpha}_1-\alpha_1}{\hat{\alpha}_1-\alpha_1 
 \hat{\alpha}_2}\, \frac{\hat{\alpha}_1-\alpha_1 \alpha_2}
 {\hat{\alpha}_1-\alpha_1 \alpha_2 \hat{\alpha}_3}, \\
 K_{2}&=&\left(1-\frac{\hat{\alpha}_1-\alpha_1}
 {\hat{\alpha}_1-\alpha_1 \hat{\alpha}_2} \right)\,
 \frac{\hat{\alpha}_2-\alpha_2}{\hat{\alpha}_2-\alpha_2 
 \hat{\alpha}_3}, \\
 K_{3}&=& 1- K_{1}-K_{2}.
\end{eqnarray}
Eq.~(\ref{eq:transition_3i}) indicates that the transition matrix for
three successive periods of intermediate competition is equivalent to a
linear combination of three transition matrices of intermediate type,
which are different from the original ones.  The values of $\beta$ for
these intermediate competitions are given by
\begin{eqnarray} 
 \beta_1 &=& \frac{1-\hat{\alpha}_1}{1-\alpha_1 \alpha_2 \alpha_3}, 
 \\
 \beta_2 &=& \frac{1-\alpha_1 \hat{\alpha}_2}{1-\alpha_1 \alpha_2 
 \alpha_3}, \\
 \beta_3 &=& \frac{1-\alpha_1 \alpha_2 \hat{\alpha}_3}{1-\alpha_1 
 \alpha_2 \alpha_3}.
\end{eqnarray}
Note that these values satisfy $\beta_1 < \beta_2 < \beta_3$, which can
be obtained from $0<\alpha_i<\hat{\alpha}_i <1$.

\section{Discussion}
\label{sect:discussion}
In this study, we have considered the effect of competition on
population dynamics in a situation that shows dependence of the types of
resource and competition on seasons or growth stages.  Introducing
transition matrices based on the number of individuals alive within a
site, we found that the effect of two successive competition periods is
summarized in simple relations on products of two transition matrices,
as shown in Table~\ref{table:products_of_matrices}.  Transition matrices
for more than two successive periods can be obtained by the recursive
use of the relations above, generally resulting in a linear combination
of various transition matrices of intermediate competition.  In
particular, transition matrices for any number of periods of either
scramble or contest result in just a single transition matrix of the
same competition type.  This suggests that the effect of competition in
such cases cannot be distinguished from that of competition involving
only a single resource.

Three comments on conditions assumed in this study are in order.  First,
in combining periods of contest or intermediate competition, the
resource at each site was assumed to be partitioned in the order of
competitive ability, but the order in each period was allowed be
independent.  Thus, we can also consider the order to be determined
randomly, i.e., not according to competitive ability, and hence in that
case, only fortunate individuals obtain the necessary amount of resource
earlier.  Second, this study considered only the case in which
individuals compete within the same site throughout the overall
competition period, but in some species, individuals may compete at
different places at different life stages.  The results of this study
cannot be applied to such situations.  Third, for simplicity, simplified
assumptions were made in this study.  For example, each individual was
alway assumed to die if the amount of resource $i$ obtained was smaller
than $s_i$ even by a small quantity. Furthermore, no differences between
individuals were taken into account, except for their competitive
ability.  How the results change when we make more realistic assumptions
is an issue left to address in future investigations.

Last, a comment on the extension of the results of this study: This
study did not consider cases in which individuals compete for multiple
kinds of resources simultaneously, but these types of cases can be
studied on the basis of the results of the present study.  We believe
that arguments based on transition matrices and the relationships
between them, obtained in this study, can be applied to various
situations for studying the effect of competition between individuals on
population dynamics.

\section*{Acknowledgments}
The author thank H. Seno for a discussion that inspired this
study.

\renewcommand{\theequation}{A.\arabic{equation}}
\setcounter{equation}{0}

\section*{Appendix A}
This appendix presents a derivation of
Eq.~(\ref{eq:transition_i_result}), which is an identity on the
following product of two transition matrices
\begin{equation}
 T_{k, m}=\sum_{l=1}^{k} T_I(\hat{\alpha}_1, \alpha_1)_{k, l}\,
 T_I(\hat{\alpha}_2, \alpha_2)_{l, m}.
\end{equation}
Because $T_{k,m}$ is expressed similarly to Eq.~(\ref{eq:T_i0}) as
\begin{equation}
 \label{eq:psi_phi}
 T_{k,m}=\phi_{k,m}-\phi_{k,m+1},
\end{equation}
we first determine $\phi_{k,m}$, which is given by
\begin{equation}
 \label{eq:phi_km}
 \phi_{k, m}=\sum_{l=1}^{k} T_I(\hat{\alpha}_1, \alpha_1)_{k, l}\,
 \phi_I(\hat{\alpha}_2, \alpha_2)_{l, m}.
\end{equation}
Here, $\phi_I(\hat{\alpha}_i, \alpha_i)_{l, m}$ is given by
\begin{equation}
 \label{eq:phi_i}
 \phi_I(\hat{\alpha}_i, \alpha_i)_{l, m}
 = \hat{\alpha}_i^l (\alpha_i/\hat{\alpha}_i)^m I_{l\ge m}
\end{equation}
from Eq.~(\ref{eq:phi_i0}), and $T_I(\hat{\alpha}_i, \alpha_i)$ is
derived from Eq.~(\ref{eq:T_i0}).  Combining these with
Eq.~(\ref{eq:phi_km}) gives
\begin{eqnarray}
 \phi_{k, m}
 &=&
 \sum_{l=m}^{k-1} \hat{\alpha}_1^k 
 \Bigl(\frac{\alpha_1 \hat{\alpha}_2}{\hat{\alpha}_1}\Bigr)^l
 \Bigl(1-\frac{\alpha_1}{\hat{\alpha}_1}\Bigr)
 \Bigl(\frac{\alpha_2}{\hat{\alpha}_2}\Bigr)^m
 \nonumber \\
 &+&
 (\alpha_1 \hat{\alpha}_2)^k \Bigl(\frac{\alpha_2}{\hat{\alpha}_2}
 \Bigr)^m I_{k\ge m}.
\end{eqnarray}
Rewriting this equation as
\begin{eqnarray}
 \phi_{k, m}
 &=&
 \frac{1-\alpha_1/\hat{\alpha}_1}{1-\alpha_1 
 \hat{\alpha}_2/\hat{\alpha}_1}
 \biggl\{
 \sum_{l=m}^{k-1} \hat{\alpha}_1^k 
 \Bigl(\frac{\alpha_1 \hat{\alpha}_2}{\hat{\alpha}_1}\Bigr)^l
 \Bigl(1-\frac{\alpha_1 \hat{\alpha}_2}{\hat{\alpha}_1}\Bigr)
 \nonumber \\
 &+&(\alpha_1 \hat{\alpha}_2)^k \biggr\}
 \Bigl(\frac{\alpha_2}{\hat{\alpha}_2}\Bigr)^m I_{k\ge m}
 \nonumber \\
 &+&\biggl(1-\frac{1-\alpha_1/\hat{\alpha}_1}
 {1-\alpha_1 \hat{\alpha}_2/\hat{\alpha}_1}\biggr)
 (\alpha_1 \hat{\alpha}_2)^k \Bigl(\frac{\alpha_2}{\hat{\alpha}_2}
 \Bigr)^m I_{k\ge m},
 \label{eq:phi_km2}
\end{eqnarray}
and performing the summation in the right-hand side gives
\begin{eqnarray}
 \phi_{k, m}
 &=& 
 \frac{\hat{\alpha}_1-\alpha_1}{\hat{\alpha}_1-\alpha_1 
 \hat{\alpha}_2}\,
 \hat{\alpha}_1^k 
 \Bigl(\frac{\alpha_1 \alpha_2}{\hat{\alpha}_1}\Bigr)^m 
 I_{k\ge m} \nonumber \\
 &&
 +\biggl(1 -\frac{\hat{\alpha}_1-\alpha_1}
 {\hat{\alpha}_1-\alpha_1 \hat{\alpha}_2}\biggr)\,
 (\alpha_1 \hat{\alpha}_2)^k \Bigl(\frac{\alpha_1 \alpha_2}
 {\alpha_1 \hat{\alpha}_2}\Bigr)^m I_{k\ge m}.
\end{eqnarray}
Further, comparing this equation with Eq.~(\ref{eq:phi_i}) renders
\begin{equation}
 \label{eq:phi_km_result}
 \phi_{k, m}= K\,\phi_I(\hat{\alpha}_1,\alpha_1\alpha_2)_{k, m}
 + (1-K)\, \phi_I(\alpha_1 \hat{\alpha}_2,
 \alpha_1\alpha_2)_{k, m},
\end{equation}
where $K=(\hat{\alpha}_1-\alpha_1)/ (\hat{\alpha}_1-\alpha_1
\hat{\alpha}_2)$.  Combining Eq.~(\ref{eq:phi_km_result}) with
Eq.~(\ref{eq:psi_phi}) finally gives
\begin{equation}
 T_{k,m} = K \,T_I(\hat{\alpha}_1, \alpha_1 \alpha_2)_{k,m}
  + (1-K)\,T_I(\alpha_1 \hat{\alpha}_2, \alpha_1\alpha_2)_{k,m},
\end{equation}
which yields the relation in Eq.~(\ref{eq:transition_i_result}).

\renewcommand{\theequation}{B.\arabic{equation}}
\setcounter{equation}{0} 

\section*{Appendix B}
This appendix privides a description of how
Eq.~(\ref{eq:transition_i_result}) is obtained from the set of
Eqs.~(\ref{eq:tra_ss})--(\ref{eq:tra_cs}).  First, with
Eq.~(\ref{eq:tra_sc}), expressing each $T_I$ in the product of two $T_I$
as a product of $T_S$ and $T_C$ gives
\begin{eqnarray}
 \lefteqn{T_I(\hat{\alpha}_1, \alpha_1) \, T_I(\hat{\alpha}_2, 
 \alpha_2)} \nonumber \\
  &=&
 T_S(\hat{\alpha}_1) \, T_C(\alpha_1/\hat{\alpha}_1) \,
 T_S(\hat{\alpha}_2) \, T_C(\alpha_2/\hat{\alpha}_2).
 \label{eq:app_b1}
\end{eqnarray}
Next, rewriting $T_C(\alpha_1/\hat{\alpha}_1) \, T_S(\hat{\alpha}_2)$ in
expression (\ref{eq:app_b1}) with Eq.~(\ref{eq:tra_cs}) shows that the
right-hand side of (\ref{eq:app_b1}) can be written as
\begin{eqnarray}
&&\!\!\!\!\!\!\!\!\!\!\!\!\!\! T_S(\hat{\alpha}_1)
 \left\{
 \frac{1-\alpha_1/\hat{\alpha}_1}{1-\alpha_1 
 \hat{\alpha}_2/\hat{\alpha}_1}
 T_C(\alpha_1 \hat{\alpha}_2/\hat{\alpha}_1) \right.
 \nonumber \\
 &+& \left. \Bigl(1-\frac{1-\alpha_1/\hat{\alpha}_1}
 {1-\alpha_1 \hat{\alpha}_2/\hat{\alpha}_1}\Bigr)
 T_S(\alpha_1 \hat{\alpha}_2/\hat{\alpha}_1)
 \right\}\, T_C(\alpha_2/\hat{\alpha}_2)
 \nonumber \\
 \label{eq:app_b2}
 &=&
 \frac{\hat{\alpha}_1-\alpha_1}{\hat{\alpha}_1-\alpha_1 
 \hat{\alpha}_2}
 T_S(\hat{\alpha}_1) T_C(\alpha_1 \hat{\alpha}_2/\hat{\alpha}_1) 
 T_C(\alpha_2/\hat{\alpha}_2)
 \nonumber \\
 &+& \Bigl(1-\frac{\hat{\alpha}_1-\alpha_1}{\hat{\alpha}_1-
 \alpha_1 \hat{\alpha}_2}\Bigr)
 T_S(\hat{\alpha}_1) T_S(\alpha_1 \hat{\alpha}_2/\hat{\alpha}_1) 
 T_C(\alpha_2/\hat{\alpha}_2).
\end{eqnarray}
Using Eqs.~(\ref{eq:tra_ss}) and (\ref{eq:tra_cc}), expression
(\ref{eq:app_b2}) can be written as
\begin{eqnarray}
 \label{eq:app_b3}
 &&\!\!\!\!\!\!\!\!\!\!\!\!\!\! 
 \frac{\hat{\alpha}_1-\alpha_1}{\hat{\alpha}_1-\alpha_1 
 \hat{\alpha}_2}
 T_S(\hat{\alpha}_1) T_C(\alpha_1 \alpha_2/\hat{\alpha}_1) 
 \nonumber \\
 &+& \Bigl(1-\frac{\hat{\alpha}_1-\alpha_1}{\hat{\alpha}_1-
 \alpha_1 \hat{\alpha}_2}\Bigr)
 T_S(\alpha_1 \hat{\alpha}_2)T_C(\alpha_2/\hat{\alpha}_2).
\end{eqnarray}
Finally, with Eq.~(\ref{eq:tra_sc}), expressing each $T_S T_C$ in
(\ref{eq:app_b3}) as $T_I$ shows that expression (\ref{eq:app_b3}) is
identical to the right-hand side of Eq.~(\ref{eq:transition_i_result}).

\renewcommand{\theequation}{C.\arabic{equation}}
\setcounter{equation}{0} 

\section*{Appendix C}
From Eq.~(\ref{eq:pop_cs_}), the slope $u$ of the reproduction curve of
the CS model at the positive equilibrium point is written as follows:
\begin{equation}
 u=\frac{b'\alpha}{(1+y/\lambda)^{\lambda+1}}
 \left\{1-\gamma_s \Bigl(1+\frac{1}{\lambda}\Bigr)
 \frac{y}{1+y/\lambda}\right\},
\end{equation}
where $y=(1-\alpha)x_*/n$, and $x_*$ is the population size at the
equilibrium point.  Since
\begin{equation}
\gamma_s \Bigl(1+\frac{1}{\lambda}\Bigr) \frac{y}{1+y/\lambda}
< \gamma_s (\lambda +1),
\end{equation}
$u$ is always positive if $1-\gamma_s(\lambda +1)>0$, and hence, the
equilibrium point is always a point attractor if $\gamma_s<1/(\lambda
+1)$.

\bibliography{ana1005_ref}

\newpage
\renewcommand{\theequation}{\arabic{equation}}

\begin{table}[htbp]
\rotatebox{90}{\begin{minipage}{\textheight}
\caption{
Interaction functions and population models for a single competition
period ($N=1$)
}
\label{table:N=1}
\begin{center}
    \small
    \renewcommand{\arraystretch}{2.2}
    \begin{tabular}{lll}
     \hline
      Type & $\phi(k)=$ & $x_{t+1}=$ \\
     \hline
      S 
      & $\phi_S(k\,; \alpha) \equiv (b' \alpha) k \alpha^{k-1}$
      & $f_S(x_t\, ; \alpha, \lambda) \equiv (b' \alpha) x_t 
        \left( 1+ \frac{(1-\alpha)x_t}{\lambda n} 
        \right)^{-\lambda-1}$ 
      \\
     C
     & $\phi_C(k\,; \alpha) \equiv (b' \alpha) 
       \frac{1-\alpha^k}{1-\alpha}$
     & $f_C(x_t\, ; \alpha, \lambda) \equiv 
       \frac{(b' \alpha)n}{1-\alpha}
       \Bigl\{1-\Bigl( 1+ \frac{(1-\alpha)x_t}{\lambda n} 
       \Bigr)^{-\lambda}\Bigr\}$
     \\
     I
     & $\phi_I(k\,; \hat{\alpha}, \alpha) \equiv (b' \alpha) 
       \frac{\hat{\alpha}^k - \alpha^k}{\hat{\alpha}-\alpha}$ 
     & $f_I(x_t\, ; \alpha, \beta, \lambda) \equiv 
       \frac{(b' \alpha)n}{(1-\alpha)(1-\beta)}  
       \Bigl\{ \left(1+ \beta \frac{(1-\alpha)x_t}{\lambda n} 
       \right)^{-\lambda}
       - \left(1+ \frac{(1-\alpha)x_t}{\lambda n} 
       \right)^{-\lambda} \Bigr\}$
     \\
     \hline
    \end{tabular}
\end{center}
\begin{center}
{\scriptsize
  S, scramble competition; C, contest competition;
  I, intermediate competition.
}
\end{center}
\end{minipage}}
\end{table}

\setcounter{equation}{\value{CNT}}
\begin{table}[htbp]
\rotatebox{90}{\begin{minipage}{\textheight}
\caption{
 Products of two transition matrices
}
\label{table:products_of_matrices}
\begin{center}
    \small
    \renewcommand{\arraystretch}{1.8}
    \begin{tabular}{lll}
     \hline
      Type & Relation \\
     \hline
      II
      & $T_I(\hat{\alpha}_1, \alpha_1) \, 
      T_I(\hat{\alpha}_2, \alpha_2)
      =\frac{\hat{\alpha}_1-\alpha_1}{\hat{\alpha}_1-\alpha_1 
      \hat{\alpha}_2}\,
      T_I(\hat{\alpha}_1, \alpha_1 \alpha_2) 
      +\frac{\alpha_1(1-\hat{\alpha}_2)}{\hat{\alpha}_1-\alpha_1 
      \hat{\alpha}_2}\,
      T_I(\alpha_1 \hat{\alpha}_2, \alpha_1 \alpha_2)$
      & $(\ref{eq:transition_i_result})$
      \\
      SS
      & $T_S(\alpha_1) \, T_S(\alpha_2) = T_S(\alpha_1 \alpha_2)$
      & $\refstepcounter{equation}(\theequation)\label{eq:tra_ss}$
      \\
      CC
      & $T_C(\alpha_1) \, T_C(\alpha_2) = T_C(\alpha_1 \alpha_2)$
      & $\refstepcounter{equation}(\theequation)\label{eq:tra_cc}$
      \\
      SC
      & $T_S(\alpha_1) \, T_C(\alpha_2) 
      = T_I(\alpha_1, \alpha_1 \alpha_2)$
      & $\refstepcounter{equation}(\theequation)\label{eq:tra_sc}$
      \\
      CS
      & $T_C(\alpha_1) \, T_S(\alpha_2)
      =\frac{1-\alpha_1}{1-\alpha_1 \alpha_2}\,T_C(\alpha_1 \alpha_2) 
      +\frac{\alpha_1(1-\alpha_2)}{1-\alpha_1 \alpha_2}\,
      T_S(\alpha_1 \alpha_2)$
      & $\refstepcounter{equation}(\theequation)\label{eq:tra_cs}$
      \\
     \hline
    \end{tabular}
\end{center}
\begin{center}
{\scriptsize
  II, intermediate followed by intermediate;
  SC, scramble followed by contest; etc.
}
\end{center}
\end{minipage}}
\end{table}

\begin{table}[htbp]
\rotatebox{90}{\begin{minipage}{\textheight}
\caption{
 Population models for two successive competition periods
}
\label{table:pop_models}
\begin{center}
    \small
    \renewcommand{\arraystretch}{1.8}
    \begin{tabular}{lll}
     \hline
      Type & $x_{t+1}=$ &  \\
     \hline
      II
      & $\frac{\hat{\alpha}_1-\alpha_1}
      {\hat{\alpha}_1-\alpha_1 \hat{\alpha}_2}\,
      f_I(x_t\,; \alpha_1 \alpha_2, \frac{1-\hat{\alpha}_1}
      {1-\alpha_1\alpha_2}, \lambda) 
      +\frac{\alpha_1(1-\hat{\alpha}_2)}
      {\hat{\alpha}_1-\alpha_1 \hat{\alpha}_2}\,
      f_I(x_t\,; \alpha_1 \alpha_2, \frac{1-\alpha_1\hat{\alpha}_2}
      {1-\alpha_1\alpha_2}, \lambda)$ 
      & $\refstepcounter{equation}(\theequation)\label{eq:pop_ii}$
      \\
      SS
      & $f_S(x_t\, ; \alpha_1 \alpha_2, \lambda)$
      & $\refstepcounter{equation}(\theequation)\label{eq:pop_ss}$
      \\
      CC
      & $f_C (x_t\, ; \alpha_1 \alpha_2, \lambda )$
      & $\refstepcounter{equation}(\theequation)\label{eq:pop_cc}$
      \\
      SC
      & $f_I (x_t\, ; \alpha_1 \alpha_2, \frac{1-\alpha_1}
      {1-\alpha_1 \alpha_2}, \lambda )$
      & $\refstepcounter{equation}(\theequation)\label{eq:pop_sc}$
      \\
      & $\frac{(b' \alpha_1 \alpha_2) x_t}
      {\left[1+(1-\alpha_1)x_t/n \right]
      \left[1+(1-\alpha_1 \alpha_2)x_t/n \right]}$ 
      \quad(for $\lambda=1$) 
      & $\refstepcounter{equation}(\theequation)\label{eq:pop_sc_lm=1}$
      \\
      CS
      & $\frac{1-\alpha_1}{1-\alpha_1 \alpha_2} f_C(x_t\, ; 
      \alpha_1 \alpha_2, \lambda)
      + \frac{\alpha_1(1-\alpha_2)}{1-\alpha_1 \alpha_2}
      f_S (x_t\, ; \alpha_1 \alpha_2, \lambda)$
      & $\refstepcounter{equation}(\theequation)\label{eq:pop_cs}$
      \\
      & $(b' \alpha_1 \alpha_2)\, x_t \,\frac{1+(1-\alpha_1)x_t/n}
      {\left[1+(1-\alpha_1 \alpha_2)x_t/n \right]^2}$ 
      \quad (for $\lambda=1$) 
      & $\refstepcounter{equation}(\theequation)\label{eq:pop_cs_lm=1}$
      \\
     \hline
    \end{tabular}
\end{center}
\begin{center}
{\scriptsize
  II, intermediate followed by intermediate;
  SC, scramble followed by contest; etc.
}
\end{center}
\end{minipage}}
\end{table}

\end{document}